\documentclass[useAMS,usenatbib]{mn2e}
\usepackage{graphicx}
\usepackage{natbib}

\newcommand{\text}[1]{\quad\mbox{#1}\quad}

\newcommand{\polind}{\gamma}
\newcommand{\lrz}{\Gamma}

\newcommand{\enth}{h}
\newcommand{\dens}{\rho}

\newcommand{\vflux}{\Psi}

\newcommand{\velpot}{\Phi_\Psi}
\newcommand{\cf}{c_{\rm f}}

\newcommand{\ggf}{f}
\newcommand{\fff}{\psi}
\newcommand{\expov}{\lambda}
\newcommand{\vvartheta}{\vartheta}
\newcommand{\thetag}{\theta_{\rm m}}

\newcommand{\apj}{ApJ}

\newcommand{\apjl}{ApJ}
\newcommand{\mnras}{MNRAS}

\newcommand{\na}{NewA}

\title[Rarefaction acceleration in GRB jets]
{Rarefaction acceleration in magnetized gamma-ray burst jets}
\author[K. Sapountzis and N. Vlahakis]
{Konstantinos Sapountzis\thanks
{E-Mail: ksapountzis@phys.uoa.gr}
and Nektarios Vlahakis\thanks{E-Mail: vlahakis@phys.uoa.gr}
\\ Department of Astrophysics, Astronomy and Mechanics,
Faculty of Physics, University of Athens,
15784 Zografos, Athens, Greece}

\begin{document}
\date{Received/Accepted}
\pagerange{\pageref{1779}--\pageref{1788}} \pubyear{2013}
\maketitle
\label{firstpage}

\begin{abstract}
Relativistic jets associated with long/soft gamma-ray bursts are
formed and initially propagate in the interior of the progenitor star.
Because of the subsequent loss of their external pressure support
after they cross the stellar surface, these flows can be modeled as
moving around a corner. A strong steady-state rarefaction wave is
formed, and the sideways expansion is accompanied by a
{\em rarefaction acceleration}.
We investigate the efficiency and the general characteristics of this
mechanism by integrating the steady-state, special relativistic,
magnetohydrodynamic equations, using a special set of partial exact
solutions in planar geometry ($r$ self-similar with respect to the
``corner''). We also derive analytical approximate scalings
in the ultrarelativistic cold/magnetized, and hydrodynamic limits.
The mechanism is more effective in magnetized than in purely
hydrodynamic flows. It substantially increases the Lorentz factor
without much affecting the opening of the jet; the resulting values
of their product can be much grater than unity, allowing for
possible breaks in the afterglow light curves. These findings are
similar to the ones from numerical simulations of axisymmetric jets by
Komissarov et al and Tchekhovskoy et al, although in our approach we
describe the rarefaction as a steady-state simple wave and
self-consistently calculate the opening of the jet that corresponds to
zero external pressure.
\end{abstract}

\begin{keywords}
gamma-ray burst: general -- MHD -- methods: analytical
-- relativistic processes
\end{keywords}

\section{Introduction}
\label{introduction}

The commonly accepted paradigm for gamma-ray bursts (GRBs) is that
they are formed in ultrarelativistic, collimated jets.  Typical
terminal Lorentz factors for these jets, such that the photons can
freely escape, are $\lrz_j \sim$ a few hundreds, even above on thousand
(e.g. \citealp{LS01,ZLB11}).
Opening angles $\Theta_j$ are inferred from achromatic breaks in the
afterglow light curves, although these are not clearly detected in
several bursts (e.g., \citealp{Liang08,Racusin09,Cenko10}).
Since the Lorentz factor decreases during the afterglow phase,
and the break in the light curves occurs when the beaming angle
equals $\Theta_j$, the product $\lrz_j\Theta_j$ should be larger than
one at the start of this phase, typically of the order of a few tens.

The long/soft class of GRBs are thought to be connected with the
death of massive stars, since some of them are associated with
Type Ic supernovae and are observed in star-forming regions of the
host galaxies (see e.g., \citealp{Zhangreview11} and references therein).
During this process a compact central object and accretion disk are formed,
and the jet is powered by either the neutrino annihilation, or by
magnetic fields, tapping the rotational energy of the central object or disk.
It is not clear which of the two mechanisms (or both) operates, with the
detection or not of the thermal photospheric emission
being a key factor \citep{ZP09,Peer12}.

Thermal (fireball) acceleration is in general a fast and efficient process,
with the Lorentz factor increasing linearly with the cylindrical distance
from the symmetry axis.
Magnetic acceleration also works provided that the jet is supported
externally by an environment whose pressure does not drop faster than
the inverse square of the distance from the origin,
as was analytically shown in \citep{KVKB09}.
The interior of the progenitor star could very well play this role.
Relativistic magnetohydrodynamic (MHD) simulations by \citep{KVKB09}
show that the efficiency of the magnetic acceleration is $\sim 50\%$ or more.
The model faces two problems though:
(1)
The jet looses its external support when it exits the progenitor star,
its motion becomes ballistic and its acceleration is practically halted.
(2)
The magnetic acceleration requires that the flow is expanded in a way
such that the separation between neighboring streamlines
increases faster than the cylindrical radius. This is achieved
through stronger collimation of the inner part of the outflow relative to the
outer part, and for this reason the mechanism was dubbed
{\em collimation-acceleration} by \cite{KVKB09}.
The resulting jets are very narrow with $\Theta_j \sim 1/\lrz_j$,
and the product $\lrz_j \Theta_j $ is close to unity before the start
of the afterglow phase, making the breaks unlikely to happen.

A solution to both problems can be given by carefully studying the dynamics
at the regime where the jet comes out from the star, and its external pressure
drops to practically zero. \cite{TNM10} simulated this transition and found
that it is accompanied by a spurt of acceleration.
\cite{KVK10} confirmed their finding numerically and interpreted it as
{\em rarefaction acceleration}. The loss of external support induces a
sideways expansion of the jet, and a strong rarefaction wave that is
driven into the flow and accelerates it.

In fact this is a powerful mechanism seen in other numerical simulations
of both, hydrodynamic and MHD flows with contact discontinuity and flow along it,
see \cite{AR06,Mizuno08,ZHK10,MMS12}.
\\
A similar mechanism was studied in \citealp{Lyutikov11,GKS11},
for the problem of an initially static magnetized plasma allowing to move
into an environment. Here we are interested for non-static cases
and their sideways expansion.

However, the analysis in all these works were based on time dependent simple waves,
while for the GRB problem under consideration it is more appropriate
to use steady-state simple waves.
In the present paper we develop a model for the
steady-state, relativistic, magnetized, rarefaction wave.

The work is a generalization of the classical steady-state,
hydrodynamic rarefaction analyzed in \cite{Landau_raref}, and
its relativistic (but again unmagnetized) counterpart by \cite{Granik82,KS84}.

In Section~\ref{seceq} we review the steady-state, special relativistic,
MHD equations in planar geometry. In Section~\ref{secrss} we develop the model,
in Section~\ref{results} we present and discuss the results
and their application to GRBs. Finally, in Section~\ref{discuss} we give a summary.

\section{Basic equations}
\label{seceq}

The system of equations of special relativistic, ideal MHD, consist of
the Ohm's law
\begin{equation}\label{ohm}
{\bmath{E}}= -\frac{\bmath{v}}{c} \times \bmath{B}\,,
\end{equation}
the Maxwell equations
\begin{equation}\label{maxwell}
\nabla \cdot \bmath{B}=0\,,  \quad
\nabla \times \bmath{E}=-\frac{1}{c}\frac{\partial \bmath B}{\partial t}\,,
\end{equation}
the mass
\begin{equation}\label{continuity}
\frac{\partial \left(\lrz \dens\right)}{\partial t}+
\nabla \cdot \left( \lrz \dens\bmath{v} \right) =0\,,
\end{equation}
momentum
\begin{eqnarray}\label{momentum}
\lrz \dens \left(\frac{\partial}{\partial t}+{\bmath{v}} \cdot \nabla \right)
\left(\enth \lrz {\bmath{v}} \right)= -\nabla p +
\qquad \nonumber \\
\frac{\nabla \cdot \bmath{E}}{4\pi}\bmath{E} +
\left(\frac{\nabla \times \bmath{B} }{4\pi}
-\frac{1}{4\pi c}\frac{\partial \bmath E}{\partial t} \right)
\times \bmath B \,,
\end{eqnarray}
and entropy
\begin{equation}\label{entropy}
\left(\frac{\partial}{\partial t}+\bmath v \cdot \nabla \right)
p = \dens c^2
\left(\frac{\partial}{\partial t}+\bmath v \cdot \nabla \right)
\enth
\end{equation}
conservation equations (e.g., \citealp{VK03a}).
\\
Here $\bmath v$ is the velocity of the outflow,
$\lrz$ is the associated Lorentz factor satisfying
\begin{equation}\label{lorentz}
\lrz^2=1+\left(\lrz\bmath v/c\right)^2 \,,
\end{equation}
$(\bmath{E}\,,\bmath{B})$
the electromagnetic field as measured in the central object's frame,
$\dens$ the rest mass density, $p$ the gas pressure, and
$\enth=w/\dens c^2$ the specific enthalpy (over $c^2$), whose expression is,
for an ideal gas with polytropic index $\polind$,
\begin{equation}\label{enthalpy}
\enth =1+\frac{\polind}{\polind-1}\frac{p}{\dens c^2} \,.
\end{equation}
The polytropic index takes the values $4/3$ or $5/3$
in the limit of ultrarelativistic or nonrelativistic temperatures,
respectively. (Any other value would imply a nonadiabatic evolution
and hence requires the incorporation of heating/cooling terms
into the entropy and momentum equations.
See also \citealp{Chiu73} for intermediate temperatures.)

By assuming steady state ($\partial / \partial t=0$) and
a planar symmetric flow, i.e., $\partial / \partial y=0$
in a system of Cartesian spatial coordinates $\left(x,y,z\right)$,
we can carry out a partial integration of the above
equations~(\ref{ohm}--\ref{enthalpy}).
It is also sufficient to simplify the analysis
by assuming that the flow lies on the poloidal plane $x-z$, $v_y=0$,
and the magnetic field in the transverse direction,
$\bmath B=B\hat y$.
As discussed in Section~\ref{results}, these are
reasonable assumptions when the model is applied to GRB outflows.
It is possible to generalize the analysis
to planar symmetric magnetized flows with
nonzero $v_y$ and poloidal magnetic field;
this will be presented in a future paper.

The continuity equation~(\ref{continuity}) for steady flows
on the $x-z$ plane yields $\lrz \dens \bmath{v}=
\nabla \times \left[\vflux(x,z) \hat y\right]$,
and thus the flow velocity can be expressed as
\begin{equation}\label{v_p-flux}
\bmath{v}=\frac{1}{\lrz \dens}\nabla \vflux \times \hat y \,.
\end{equation}
The stream function $\vflux$ is constant along each streamline
(since $\bmath{v}\cdot \nabla \vflux =0$), and can be used
as its ``label''.

Using Ohm's equation~(\ref{ohm}) we express the electric field as
$ \bmath{E}=\left({B}/{\lrz \dens c}\right) \nabla \vflux$.
Substituting in Faraday's equation~(\ref{maxwell})
we find $\nabla \left(B/\lrz \dens c\right)\times
\nabla \vflux =0$, which means that the quantity
$B/\lrz \dens c$ is a streamline constant,
\begin{equation}\label{phipsi}
-\frac{B}{\lrz \dens c} = \velpot(\vflux) \,.
\end{equation}
Thus, the electric field can be written
as $\bmath{E}=-\velpot \nabla \vflux
=-\nabla \int \velpot d \vflux $.
This expression shows the relation of the function $\velpot$
with the scalar electric potential, and also
that the streamlines are equipotentials.

The component of the momentum equation~(\ref{momentum})
along the flow gives after some manipulation\footnote{
We apply the identity $(\bmath G \cdot \nabla)\bmath G
=  \nabla (G^2/2) +( \nabla \times\bmath G) \times\bmath G$
for $\bmath G = \enth \lrz\bmath v$
in the left-hand side of the momentum equation (\ref{momentum})
and then, by dotting with $\bmath v$, we get
$ (\dens/\enth)\bmath v \cdot \nabla (\enth^2 \lrz^2 v^2/2) +\bmath v \cdot \nabla p
=\bmath v \cdot \left[(\nabla \times \bmath B) \times\bmath B\right] /4\pi $.
The left-hand side, by replacing $\lrz^2 v^2 / c^2= \lrz^2 -1$
and using equation~(\ref{entropy}), becomes
$\lrz \dens\bmath v \cdot \nabla (\enth \lrz c^2) $.
The right-hand side,
using $\nabla \times\bmath B = \nabla B \times\hat y$
and equation~(\ref{phipsi}), can be written as
$(c/4\pi) \velpot \lrz \dens\bmath v \cdot \nabla B$.
Since $\bmath v \cdot \nabla \velpot =0$, this is equal to
$(c/4\pi) \lrz \dens\bmath v \cdot \nabla ( \velpot B)$,
and the equation of the two sides gives
$\bmath v \cdot \nabla \left[\enth \lrz c^2
-(c/4\pi) \velpot B\right]=0 $.
Thus, the quantity inside the brackets is a streamline constant.  }
\begin{equation}\label{mu}
\enth \lrz - \frac{\velpot B}{4 \pi c} =\mu(\vflux)\,.
\end{equation}
This integral represents the total energy-to-mass flux ratio
(over $c^2$),
since the mass flux (times $c^2$) is $c^2 \lrz \dens\bmath v$,
the Poynting flux $(c/4\pi)\bmath E \times \bmath B=
-({\velpot B}/{4 \pi c}) c^2 \lrz \dens\bmath v$,
and the matter energy flux (including thermal, bulk kinetic and rest
energy) is $c^2 \lrz^2 \enth \dens\bmath v$.

The entropy conservation equation~(\ref{entropy}),
for an ideal gas whose enthalpy is given by equation~(\ref{enthalpy}),
simplifies to
$\bmath{v}\cdot\nabla(p/\dens^{\polind})=0$,
meaning that the quantity $p/\dens^{\polind}$
-- which is related to the specific entropy -- is
a streamline constant
\begin{equation}\label{Q}
\frac{p}{\dens^{\polind}}=Q(\vflux) \,.
\end{equation}

The previous partial integrations greatly simplify the original
system of equations~(\ref{ohm})--(\ref{enthalpy}),
yielding several streamline constants, which can be determined
at the boundary of the flow.
Three equations remain to be integrated: the component of the
momentum equation~(\ref{momentum}) normal to the flow velocity,
and equations~(\ref{lorentz}), (\ref{enthalpy}). There are
correspondingly three unknown functions, which we choose to be the
stream function $\vflux$, the specific enthalpy $h$, and
the ratio of Poynting-to-matter energy flux
\begin{equation}\label{sigma-v}
\sigma \equiv
\frac{B^2}{4\pi \enth \lrz^2 \dens c^2}=
\frac{\velpot^2 \dens}{4 \pi \enth}\,.
\end{equation}

We may write the physical quantities in terms of these variables:
\begin{equation}\label{rho_P-v}
\dens=\frac{4 \pi \enth \sigma}{\velpot^2} \,, \quad
p=Q \dens^{\polind} \,,
\end{equation}
\begin{equation}\label{A22-v}
\lrz=\frac{\mu}{\enth \left(1+\sigma\right)}\,, \quad
\bmath{v}=\frac{\nabla \vflux \times \hat y }{\lrz \dens} \,,
\end{equation}
\begin{equation}\label{A1-v}
{\bmath {B}}=-\frac{4\pi\mu c}{\velpot}\frac{\sigma}{1+\sigma}
\hat y\,, \quad
{\bmath{E}}=-\velpot \nabla \vflux\,.
\end{equation}

Knowing the streamline constants ($\velpot$, $\mu$, $Q$)
we can find the remaining unknowns $\vflux\,, \sigma\,, \enth$
by solving the following system of equations,
with the first coming from equation~(\ref{enthalpy}),
the second from equation~(\ref{lorentz}),
and the third from the component of the
momentum equation~(\ref{momentum}) normal to the flow velocity:
\begin{equation}\label{x-M-v}
\enth=1+\frac{\polind}{\polind-1}\frac{Q}{c^2}
\left(\frac{4 \pi}{\velpot^2}\right)^{\polind-1}
(\enth \sigma)^{\polind-1} \,,
\end{equation}
\begin{equation}\label{bernoulli-v}
\frac{\mu^2}{\enth^2(1+\sigma)^2}=1+
\left(\frac{\velpot^2 \nabla \vflux}{4 \pi c \enth \sigma} \right)^2 \,,
\end{equation}
\begin{eqnarray}\label{transfield-v}
\frac{\velpot^2}{\sigma}
\left[\left(1+\sigma\right) \nabla^2 \vflux - \nabla \vflux \cdot
\nabla \ln \mid \nabla \vflux \mid \right]
\nonumber  \\
-\frac{1}{2} \nabla \left( \frac{4\pi\mu c}{\velpot}\frac{\sigma}{1+\sigma}
\right)^2 \cdot \frac{\nabla \vflux } {\mid \nabla \vflux \mid^2 }
+\frac{\mid \nabla \vflux \mid^2}{2} \frac{d\velpot^2}{d\vflux}
\nonumber  \\
-\frac{\polind-1}{\polind} \nabla \left[
16 \pi^2 c^2 \frac{\enth (\enth-1) \sigma }{\velpot^2}
\right] \cdot \frac{\nabla \vflux } {\mid \nabla \vflux \mid^2 }
=0\,.
\end{eqnarray}

\section{The \MakeLowercase{$r$} self-similar model}\label{secrss}
\begin{figure}
\centerline{\includegraphics[scale=.45]{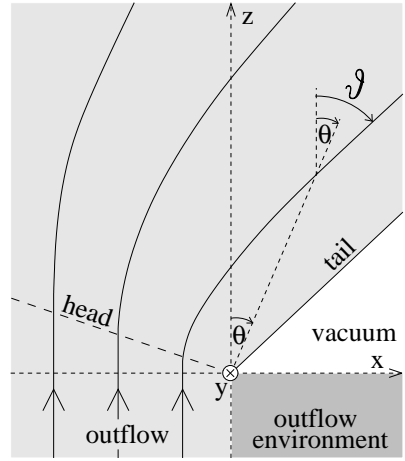}}
\caption{The outflow geometry.}
\label{sketch}
\end{figure}

The problem under consideration is basically a Prandtl-Meyer flow around a corner,
and thus the appropriate coordinates are polar on the plane $x-z$
with the corner at the origin, defined through $x=r\sin\theta$ and $z=r\cos\theta$,
see Fig.~\ref{sketch}.
A self-similar flow is described with a stream function of the form
$\vflux=r^\expov \fff(\theta) $, with constant $\expov$.
It is more convenient to replace the function $\fff(\theta)$ in terms
of the function $\ggf(\theta)$, defined through
$\fff=\vflux_0(r_0\ggf)^{-\expov}$, with constant
$\vflux_0$, $r_0$, in which case the definition of $\vflux$ yields
\begin{equation}
r=r_\vflux \ggf(\theta)\,, \quad r_\vflux \equiv
r_0\left(\frac{\vflux}{\vflux_0}\right)^{1/\expov} \,.
\end{equation}
The function $\ggf(\theta)$ gives the radial distance from the corner,
modulo a scale factor which is different in each streamline.
This clearly shows that all streamlines are similar to each-other,
hence the term ``self-similarity''.

The derivative of $\ggf(\theta)$ controls $\nabla \vflux$,
and is thus related to the flow direction.
Rewriting equation~(\ref{A22-v}) as
$\lrz \dens \bmath v = \displaystyle\frac{1}{r}\frac{\partial \vflux }{\partial \theta}
\hat r - \displaystyle\frac{\partial \vflux }{\partial r} \hat \theta$,
where $\hat r$ and $\hat \theta$ the unit vectors of the polar coordinates,
and defining the angle $\vvartheta$ between the flow velocity and the $z$ axis
(see Fig.~\ref{sketch}), the $\tan \left(\vvartheta-\theta\right)= v_\theta /v_r$ yields
\begin{eqnarray}
\frac{d\ggf}{d\theta} =\frac{\ggf}{\tan \left(\vvartheta-\theta\right)} \,.
\label{ode10}
\end{eqnarray}

Our goal is to separate the variables $r$ and $\theta$ in the
system of equations~(\ref{x-M-v})--(\ref{transfield-v}), and reduce them to
equations with respect to the polar angle $\theta$ alone.

From inspection of equations~(\ref{x-M-v}), (\ref{bernoulli-v})
we require
\begin{equation}
\sigma=\sigma(\theta) \,, \quad \enth=\enth(\theta) \,,
\mbox{ and constant } \mu \,, \ Q/\velpot^{2(\polind-1)}\,. \end{equation}
The last term of equation~(\ref{bernoulli-v})
should be a function of $\theta$ alone, and this gives the form of
the streamline constant
\begin{equation}
\velpot^2 = \frac{4 \pi c r_0}{\expov \mid\vflux_0\mid}
\left(\frac{\vflux}{\vflux_0}\right)^{\frac{1}{\expov}-1} \,.
\end{equation}
The so-called Bernoulli equation~(\ref{bernoulli-v}) can then be written as
\begin{eqnarray}
\frac{\mu^2}{\enth^2(1+\sigma)^2}=1+
\frac{1}{\enth^2 \sigma^2 \ggf^2 \sin^2\left(\vvartheta-\theta\right)} \,,
\label{odethe0}
\end{eqnarray}
or in differential form
\begin{eqnarray}
\frac{d\vvartheta}{d\theta}=-\frac{\tan\left(\vvartheta-\theta\right)
\left[\mu^2 - \enth^2 (1+u_s^2) \left(1+\sigma\right)^3\right]}
{\sigma \left(1+\sigma\right) \left[\mu^2 - \enth^2 \left(1+\sigma\right)^2\right]}
\frac{d\sigma}{d\theta} \,,
\label{odevartheta} \end{eqnarray}
where we used the differential form of equation~(\ref{x-M-v})
\begin{eqnarray}
\frac{d\enth}{d\theta}=\frac{\enth u_s^2}{\sigma}\frac{d\sigma}{d\theta}
\,, \quad u_s^2= \frac{\left(\polind-1\right)\left(\enth-1\right)}
{\polind-1+\left(2-\polind\right)\enth} \,.
\label{odeenth}\end{eqnarray}
$u_s^2$ is the square of the proper sound speed (over $c^2$)
\begin{eqnarray}
u_s^2=\frac{c_s^2/c^2}{1-c_s^2/c^2}\,, \quad
c_s^2 = \frac{\polind p}{\dens \enth} \,.
\end{eqnarray}

The transfield equation~(\ref{transfield-v}), after some manipulation
using the previous two equations, gives
\begin{eqnarray}\label{ode2}
\frac{d\sigma}{d\theta}
=-\frac{\left(\expov-1\right) \sigma }{2 \tan\left(\vvartheta-\theta\right)}
\frac{{\cal N}}{{\cal D}}\,,
\\
{\cal N}= \sigma +\frac{2}{\polind}\frac{u_s^2}{1+u_s^2}
\,, \quad
{\cal D}=\frac{1}{\enth^2 \sigma^2 \ggf^2}-\sigma\left(1+u_s^2\right)-u_s^2 \,.
\nonumber
\end{eqnarray}

After solving the system of equations~(\ref{x-M-v}), (\ref{ode10}),
(\ref{odethe0}), (\ref{ode2})
for the functions $\ggf(\theta)$, $\vvartheta(\theta)$,
$\enth(\theta)$, $\sigma(\theta)$, the physical quantities can be recovered using
\begin{eqnarray}
\label{densss}
\dens=\frac{\expov|\vflux_0|}{cr_0} \enth \sigma
\left(\frac{r}{r_0\ggf}\right)^{\expov-1}\!\! ,
\quad
\frac{p}{\dens c^2}=\frac{\left(\polind-1\right)\left(\enth-1\right)}{\polind} \,,
\\
\lrz \frac{\bmath{v}}{c}=-\frac{\vflux_0}{|\vflux_0|}
\frac{\cos\left(\vvartheta-\theta\right) \hat r + \sin\left(\vvartheta-\theta\right) \hat \theta}
{\enth \ggf \sigma\sin\left(\vvartheta-\theta\right)} \,,
\\
\lrz=\frac{\mu}{\enth \left(1+\sigma\right)}\,, \quad
\frac{\bmath {B}}{\sqrt{4\pi \dens \enth c^2}}=-\frac{\velpot}{|\velpot|}
\frac{\mu \ \sigma^{1/2}}{\enth (1+\sigma)}
\hat y\,,
\\
\frac{\bmath{E}}{\sqrt{4\pi \dens \enth c^2}} =
-\frac{\vflux_0}{|\vflux_0|} \frac{\velpot}{|\velpot|}
\frac{\sin\left(\vvartheta-\theta\right) \hat r -\cos\left(\vvartheta-\theta\right) \hat \theta
}{\enth \ggf \sigma^{1/2}\sin\left(\vvartheta-\theta\right)}
\,.
\label{ess}
\end{eqnarray}

Note that, using the previous expressions, the numerator and denominator
of the differential equation~(\ref{ode2}) can be written as
\begin{eqnarray}
{\cal N}=\frac{B^2-E^2}{4\pi \dens \enth c^2}
+ \frac{2}{\polind} \frac{u_s^2}{1+u_s^2}
\,, \nonumber \\
{\cal D}= \left(\frac{\lrz v_\theta}{c}\right)^2-
{\displaystyle \frac{B^2 - E^2}{4 \pi \enth \dens c^2} (1+u_s^2) -u_s^2}
\,.
\label{expd}
\end{eqnarray}
${\cal N}$ is always positive, while ${\cal D}$ can be written as
$(\lrz v_\theta/c)^2 - u_{\rm f}^2$
(using expression~\ref{omegacoproper} of Appendix~\ref{appendixa}).

\subsection{The rarefaction wave $(\expov=1)$ case}
\label{sec_raref}

Near the corner the flow properties are expected to depend mostly on the
polar angle $\theta$; their dependence on the coordinate $r$ is only weak.
This requires the parameter $\expov$ to be $\approx 1$,
see equations~(\ref{densss})--(\ref{ess}) (the density is
proportional to $r^{\expov-1}$ and the other
quantities depend on $r$ through the density).

The case $\expov=1$ corresponds to the classical rarefaction wave,
a steady-state simple wave. It is the
relativistic MHD generalization of the hydrodynamic steady-state
rarefaction wave analyzed by \cite{Landau_raref} in the nonrelativistic regime,
and \cite{Granik82,KS84} in the relativistic case.
As in these studies, the assumption that the flow depends only on the polar angle
$\theta$ leads to two possibilities:
the first corresponds to a uniform flow, and the second to a rarefaction wave.
By inspection of equation~(\ref{ode2}) for $\expov=1$ one directly concludes
that ${\cal D} \ d\sigma/d\theta=0$.
The case with constant $\sigma$ is the trivial one of a uniform flow,\footnote{
For constant $\sigma$ equation~(\ref{x-M-v}) implies that $\enth$ is also constant,
equation~(\ref{odethe0}) yields that $ \ggf\propto 1/\sin(\vvartheta-\theta)$,
and the combination of the latter with equation~(\ref{ode10})
gives that $\vvartheta$ is also constant.}
while ${\cal D}= 0$ corresponds to the rarefaction wave.

\begin{figure}
\centerline{\includegraphics[scale=.45]{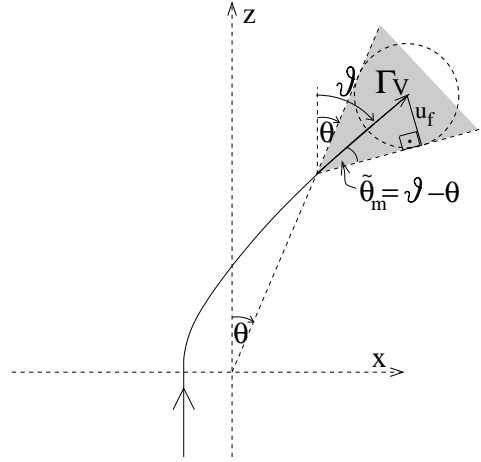}}
\caption{The Mach cone (shadowed area) with opening half-angle
$ \tilde \thetag =\vvartheta-\theta $.
The $\lrz v_\theta=\lrz v \sin \tilde \thetag$
component of the flow proper velocity equals $u_{\rm f}$ (radius of cycle).}
\label{sketch2}
\end{figure}

A more robust perspective is to notice that ${\cal D}= 0$ implies that
the $\hat \theta$ component of the flow proper velocity ($\lrz v_\theta/c $)
is equal to the comoving proper phase velocity of a magnetosonic wave $u_{\rm f}$.
Equivalently, the lines $\theta=$ constant intersect the streamlines at every point
at the Mach angle $\tilde \thetag$, i.e., $ \tilde \thetag =\vvartheta-\theta $,
see Fig.~\ref{sketch2}.
This can be seen by noting that $\lrz v_\theta /c =(\lrz v/c) \sin(\vvartheta-\theta)$
and $u_{\rm f}=(\lrz v/c) \sin\tilde \thetag $
(see equation~\ref{mach_opening} of Appendix~\ref{appendixa}).
\\
The two sides of the Mach cone are the two characteristics, with equations
\begin{eqnarray}
\frac{dx}{dz}=\tan\left(\vvartheta\pm \tilde \thetag\right)
\,, \ \mbox{ or, }\ \
\frac{rd\theta}{dr}=\tan\left(\vvartheta-\theta
\pm \tilde \thetag\right) \,.
\end{eqnarray}
For $ \tilde \thetag =\vvartheta-\theta $ we again conclude
that the minus characteristics are the cones $\theta=$ constant.

Suppose we are interested to model an outflow approaching a corner,
see Fig.~\ref{sketch}. The flow is initially uniform,
in pressure equilibrium with its environment (the $z<0$ regime in Fig.~\ref{sketch}),
and superfast-magnetosonic (the bulk velocity is higher
than the fast-magnetosonic wave speed).
In such a flow the information is propagating in a Mach cone around the flow speed,
formed by the plus and minus characteristics.
The effect of the corner is propagated
only downstream of the minus characteristic that leaves the corner,
and corresponds to the head of the rarefaction shown in Fig.~\ref{sketch}.
Each fluid parcel keeps moving with constant speed till it crosses this line.
From this point on the streamlines start to bent, the flow expands and its density,
thermal/magnetic energy flux decline. As a result of the
energy conservation the flow is accelerated.
This mechanism of converting thermal/magnetic energy into kinetic energy
of bulk motion is dubbed rarefaction acceleration by \cite{KVK10}.
The bending of streamlines and the acceleration of the flow continues
till the angle $\theta=\theta_t$, the tail of the rarefaction,
where the flow becomes ballistic and pressureless (in equilibrium with the vacuum).

Mathematically, in the initial uniform superfast-magnetosonic
part of the outflow $(-\pi< \theta < -\pi/2)$, ${\cal D}$ is positive.
The same is true in the first portion of the $\theta >-\pi/2$
regime. However, as the flow moves in that part
(see Fig.~\ref{sketch})
the $\hat\theta$ component of the flow velocity decreases
leading to a decreasing ${\cal D}$ (see equation~\ref{expd}).
Eventually ${\cal D}$ becomes zero at an angle $\theta=\theta_h$
corresponding to the head of the rarefaction, and remains zero
in the whole rarefaction phase
(for $\theta_h \le \theta\le \theta_t$).
The system of equations~(\ref{x-M-v}), (\ref{ode10}), (\ref{odethe0}), (\ref{ode2})
gives $\ggf$, $\vvartheta$, $\enth$, $\sigma$ at each
$\theta \in \left[\theta_h\,,\theta_t\right]$.
These expressions, together with their simplified versions in the limits
of cold and unmagnetized flows, are given in Appendix~\ref{appendixb}.

\subsection{The $\expov > 1$ case}

The $\expov=1$ is the most important case (and the only one with finite
density at the corner), but we kept the analysis more general
including $\expov > 1$ cases (for $\expov <1$ the density becomes
infinity at the origin).
In these cases the flow is nonuniform initially, with the density increasing
with the distance from the corner. As a result, denser parts tend to move towards the
less dense regions, and the resulting flow expansion
provides an additional acceleration mechanism on top of the rarefaction acceleration
which is still present.

\section{Results -- Application to GRB jets}
\label{results}

\begin{figure*}
\includegraphics[width=43mm,angle=0]{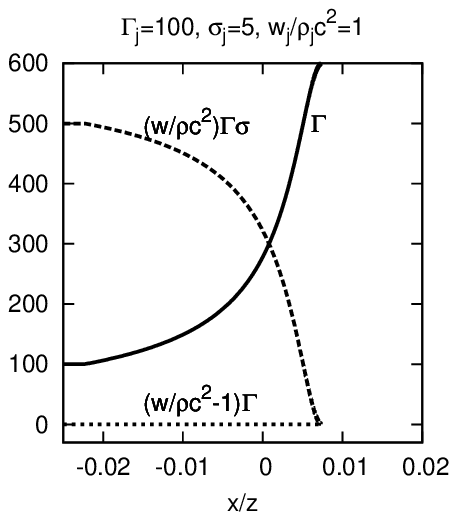}
\includegraphics[width=43mm,angle=0]{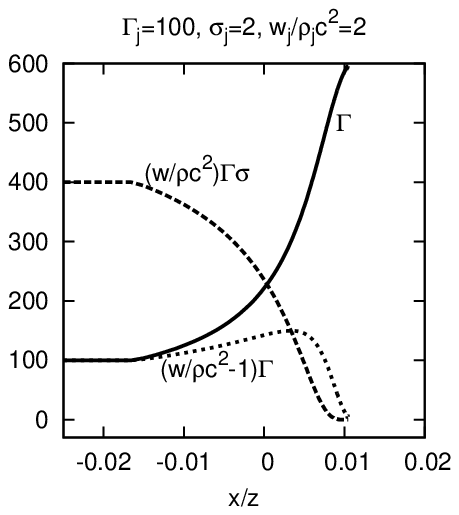}
\includegraphics[width=43mm,angle=0]{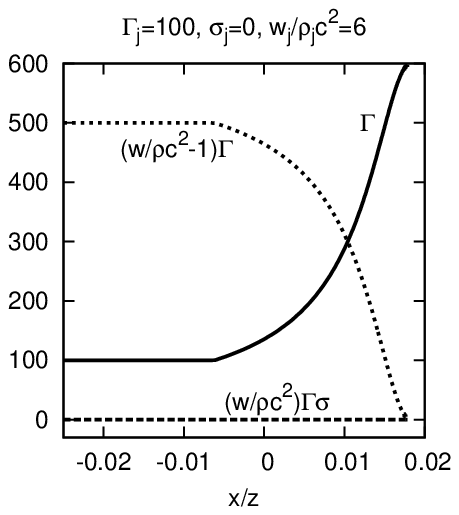}
\includegraphics[width=43mm,angle=0]{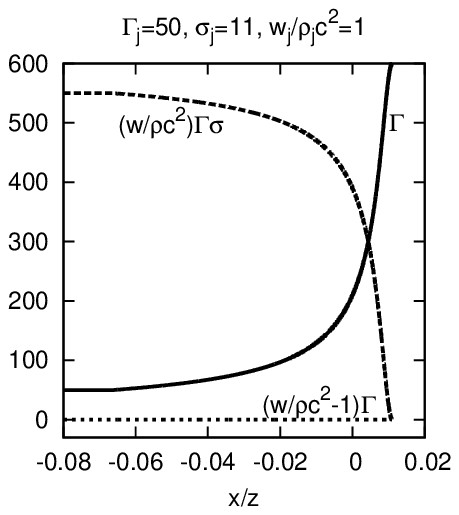}
\\
\includegraphics[width=43mm,angle=0]{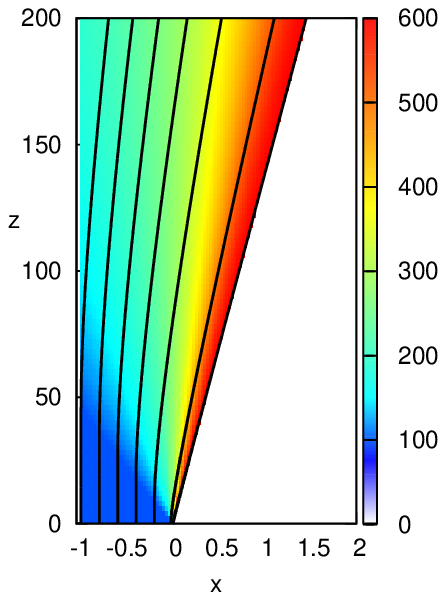}
\includegraphics[width=43mm,angle=0]{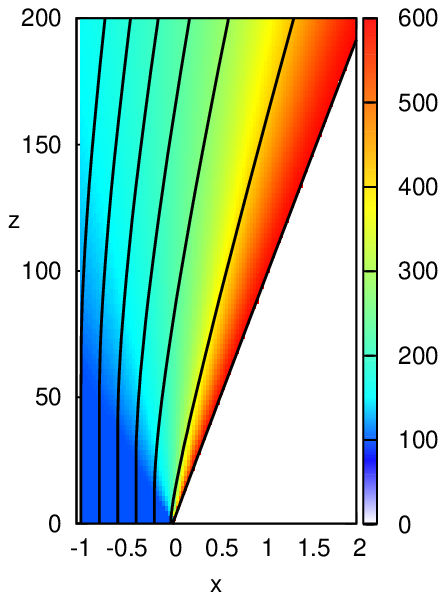}
\includegraphics[width=43mm,angle=0]{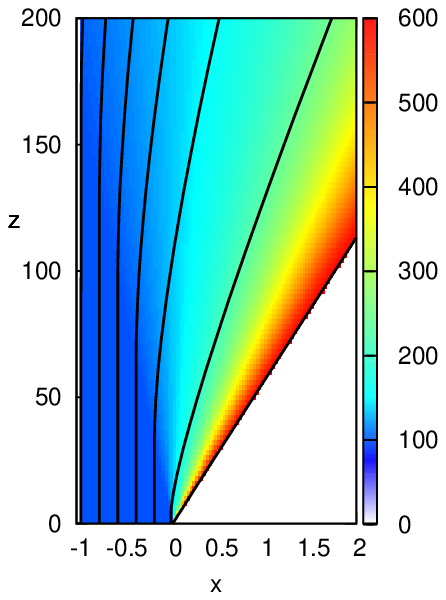}
\includegraphics[width=43mm,angle=0]{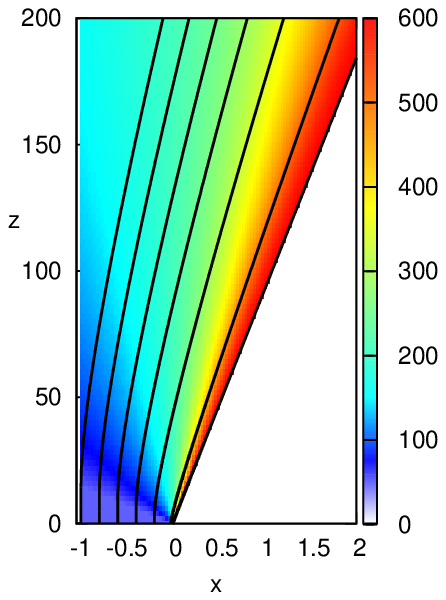}
\caption{Solutions for the rarefaction wave $\expov=1$.
The last row shows the distribution of the Lorentz factor (color) and
the streamlines (solid lines).
All cases correspond to energy-to-mass flux ratio $\mu=600$.
The scale of the distances is arbitrary. A convenient choice for the unit of distances
is the jet radius $\sim R_\star/\lrz_j$, in which case
the $x=-1$ line mimics the rotation axis of the jet.}
\label{raref_gammacont}
\end{figure*}

The numerical procedure is to give the model parameters
$\expov$, $\polind$, the initial quantities
$\lrz_j$, $\sigma_j$, $\enth_j$, and $\vvartheta_j$
at some initial angle $\theta_j$, and find $\mu$ and $\ggf_j$
using equations~(\ref{A22-v}), (\ref{odethe0}). Then
solve the system of the two algebraic equations~(\ref{x-M-v}), (\ref{odethe0})
together with the two differential equations (\ref{ode10}), (\ref{ode2})
for the functions $\ggf(\theta)$, $\vvartheta(\theta)$,
$\enth(\theta)$, $\sigma(\theta)$.

Since we are interested to apply the model to GRB outflows we set
the energy-to-mass flux ratio ($\mu$), which equals
the maximum possible bulk Lorentz factor if all the energy is transferred
to kinetic, a few hundreds. In particular,
we choose a value $\mu=600$ in the numerical results.

A jet associated with a long/soft GRB is thought to be
formed inside the progenitor star, and its first acceleration phase
takes place before it crosses the stellar surface.
We take as a reference value for the resulting bulk Lorentz factor,
which is the initial value for the rarefaction acceleration phase that
we examine, $\lrz_j=100$.
For a cold flow ($\enth_j=0$) the magnetization is $\sigma_j=5$
such that equation~(\ref{A22-v}) is satisfied.
Since the details of the acceleration phase inside
the star are not known in general,\footnote{
If the acceleration has magnetic origin, the spatial dependence of the Lorentz
factor can be approximated as $\lrz\approx (R/r_{\rm lc})^{(b-1)/b}$
where $R$ the distance from the origin and
$b$ is related to the flow shape, see \cite{KVK10}. For example,
for $b=2$ we get $\lrz_j=100 \left(\frac{R_\star}{R_\odot}\right)^{1/2}
\left(\frac{r_{\rm lc}}{5\times 10^6 {\rm cm}}\right)^{-1/2} $,
where $R_\star$ is the stellar radius,
while for $b=3/2$ we get
$\lrz_j=50 \left(\frac{R_\star}{10 R_\odot}\right)^{1/3}
\left(\frac{r_{\rm lc}}{5\times 10^6 {\rm cm}}\right)^{-1/3}$.}
we also examine a model with $\lrz_j=50$ (and $\sigma_j=11$).

If the jet is magnetically driven, it is superfast-magnetosonic
when it crosses the stellar surface.  It is well known from the MHD theory that
in this regime the magnetic field is predominantly azimuthal, justifying our
choice for ignoring the $B_x$ and $B_z$ components in the model.\footnote{
Well outside the light cylinder and for relativistic bulk motion,
the ratio of the azimuthal over the poloidal magnetic field component
equals the cylindrical distance in units of the light cylinder radius
(see, e.g., equation [33] in \citealp{KVKB09}).
For typical values of a cylindrical distance $R_\star/\lrz_j
\sim 10 R_\odot/100$ and $r_{\rm lc}=5\times 10^6$ cm this
ratio is $\sim 10^3$.}
By adopting a planar geometry we ignore the tension of the azimuthal
magnetic field. This is reasonable, since the fast variations induced by
the rarefaction wave give a much larger magnetic pressure gradient
in the radial ($x$) direction.

We also include a purely hydrodynamic model with $\sigma_j=0$
and $\enth_j=6$ (from equation~[\ref{A22-v}]),
and an intermediate case with $\sigma_j=2$ and $\enth_j=2$.

In all cases we started the integration from $\theta_j=-\pi/2$,
with a flow parallel to the $z$ axis, $\vvartheta_j=0$.

The results of the numerical integration for the rarefaction
case $\lambda=1$ are shown in Fig.~\ref{raref_gammacont}
for various sets of the initial quantities
$\lrz_j$, $\sigma_j$ and $\enth_j\equiv w_j/\rho_j c^2$.
The first column corresponds to the cold/magnetized case,
the third to the hydrodynamic/unmagnetized, and the middle
to the intermediate case.
In each column the top panels show the variation of the three
parts of the energy-to-mass flux ratio (whose sum is the constant $\mu$):
the bulk kinetic (including the rest mass energy) which is the Lorentz factor,
the Poynting which is written through the magnetization as $\enth \lrz \sigma$,
and the enthalpy $(\enth-1)\lrz$. During the rarefaction phase
the bulk acceleration to its full completion ($\lrz =\mu$) is clearly seen.

The bottom panels show the geometry of the flow (the solid lines are streamlines),
together with the Lorentz factor (color).
In agreement with the discussion in Section~\ref{sec_raref}, three distinct regimes
can be observed. The first is the unperturbed flow region
($-\pi/2 \le \theta \le \theta_h$) where $\lrz=\lrz_j$.
The second is the rarefied region ($\theta_h \le \theta \le \theta_t$) where $\lrz$ increases.
The perturbed region does not fill the whole space; there is a maximum angle $\theta_t$ --
the so-called Prandtl-Meyer angle, or the tail of the rarefaction --
leaving the rest of the area ($\theta >\theta_t$) void.

The pressure equilibrium at the contact discontinuity between the flow and the void space
($\theta=\theta_t$) implies that the thermal and magnetic pressures vanish.
Consequently the flow is ballistic along the streamline that pass through the corner,
and the whole energy flux has been already transferred to kinetic energy flux
($\lrz=\mu$).
All other streamlines are starting to bent when they cross the head of the rarefaction
and asymptotically they become parallel to the tail.
During this phase the flow is accelerating, reaching $\lrz=\mu$ asymptotically.
The spatial scale in which this acceleration takes place
strongly depends on the magnetization of the
flow, something that has important consequences for the applications of the model.
The bottom panels of Fig.~\ref{raref_gammacont} show that the cold/magnetized case
(first column) is accelerated much faster compared to the hydrodynamic case (third column),
with the intermediate case (second column) lying between these two limiting cases as
expected. For example, when the streamline starting from $x_i=-0.02$ reaches
$z=200$ it has already $\lrz \approx \mu$ in the former case, while in the later
$\lrz <\mu/2$.

\begin{figure}
\includegraphics[width=41mm,angle=0]{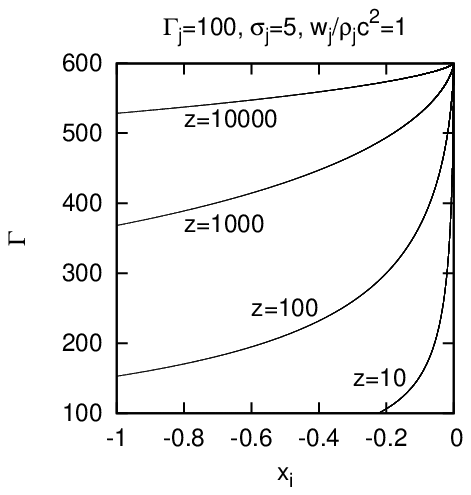}
\includegraphics[width=41mm,angle=0]{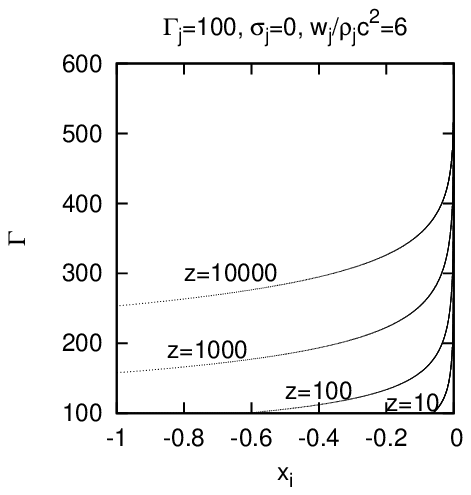}
\caption{The Lorentz factor as function of the starting position of each
fluid parcel on the $x$ axis ($x_i$), for various $z$, and for two models:
a cold/magnetized case (left) and a hydrodynamic case (right).
The scale of the distances is arbitrary.}
\label{lraref_energy}
\end{figure}
\begin{figure}
\includegraphics[width=41mm,angle=0]{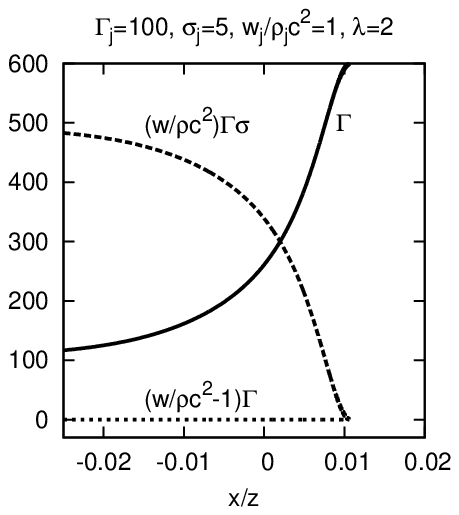}
\includegraphics[width=41mm,angle=0]{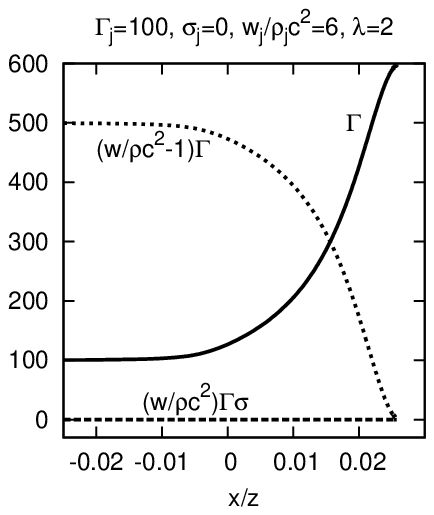}
\\
\includegraphics[width=41mm,angle=0]{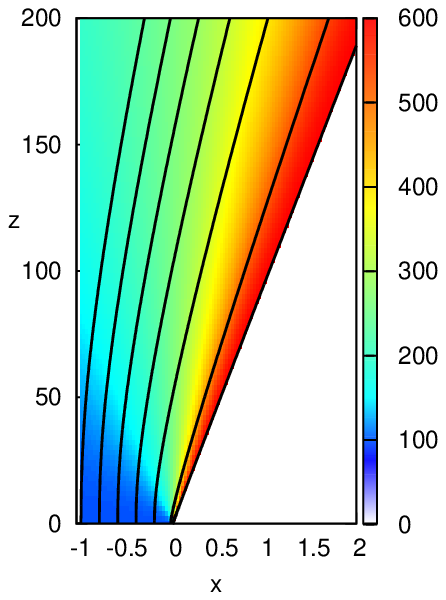}
\includegraphics[width=41mm,angle=0]{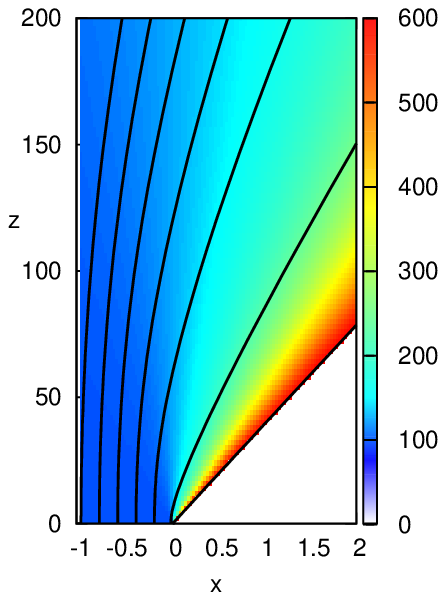}
\caption{Similarly to Fig.~\ref{raref_gammacont}, but for two solutions with $\expov=2$.}
\label{lraref_gammacont}
\end{figure}

The numerical results are in a perfect agreement with the analytical relations given in
Appendix~\ref{appendixb} and summarized below.
For the cold/magnetized case which is the most important
and most efficient, the head of the rarefaction wave is located at
$\theta_h=-\sigma_j^{1/2}/\lrz_j$ (corresponding to the
half-opening angle of the Mach cone
for the fast-magnetosonic waves, see Appendix~\ref{appendixa}).
The tail is located at $\theta_t=2\sigma_j^{1/2}/\lrz_j(1+\sigma_j)$.
Note that this angle is always smaller than $1/\lrz_j$.
If the flow inside the progenitor star is magnetically accelerated then
its half-opening angle is expected to be $1/\lrz_j$ \citep{KVKB09}.
Since $\theta_t<1/\lrz_j$, the rarefaction increases the Lorentz factor
without affecting much the opening angle, meaning that the product of the
Lorentz factor with the half-opening angle increases up to the
value $\sim \mu/\lrz_j$ when the Lorentz factor attains its maximum value $\mu$.
\\
As shown in the Appendix~\ref{appendixb}
during the acceleration the magnetization decreases as $\sigma =f^{-2/3}$,
where $f$ is proportional to the distance from the corner.
A streamline starting at $x_i$ on the $x$ axis crosses the head of the rarefaction
at $r_i=x_i/\theta_h$. Thus, $\sigma/\sigma_j=(r \theta_h/x_i)^{-2/3} $
and we get an analytical approximate expression for the Lorentz factor
\begin{equation}\label{gamma-r}
\lrz= \frac{\mu}{1+\left(\sigma_j \lrz_j x_i /r\right)^{2/3}} \,.
\end{equation}
The distance $|x_i|$ spans a range from zero -- corresponding to the corner --
up to a maximum value corresponding to the distance between the corner and
the rotation axis of the jet, i.e., the jet radius,
which can be approximated as $R_\star/\lrz_j$.
At distance $r=\sigma_j \lrz_j |x_i|$ from the corner along each streamline
(i.e., for each $x_i$), the Lorentz factor reaches half of its maximum value.
As expected, fluid parcels on streamlines that are closer to the corner
accelerate faster.
\\
In terms of the streamline shape, equation~(\ref{B10}) gives
the analytical approximate relation between the Lorentz factor and the
angle $\vvartheta$ between the flow speed and its initial orientation,
\begin{equation}
\lrz= \frac{\mu}{1+\sigma_j\left(1-\vvartheta/\theta_t\right)^{2}} \,.
\end{equation}

In the hydrodynamic case the angle $|\theta_h|$ is smaller because
the sound speed is smaller compared to the fast-magnetosonic speed.
As a result the acceleration phase starts later and needs larger
distances to reach completion.
During the acceleration a combination of equations~(\ref{B11}) and
(\ref{B13}) gives the Lorentz factor as a function of $r$.
For $\polind=4/3$ and $\lrz \ge \mu/2$ we get the approximate result
$r\propto \varrho^{-7/6}$ and thus
$\lrz = \frac{\mu}{1+{\cal C} (x_i/r)^{2/7}}$, with constant ${\cal C}$.
From this expression it is evident that the acceleration is much slower
compared to the magnetized case.

Fig.~\ref{lraref_energy} shows the result of the acceleration across the jet,
for two models. Clearly the cold/magnetized case (left panel) is much
faster accelerated compared to the hydrodynamic case (right panel).
Choosing the radius of the jet $(\sim R_\star/\lrz_j)$ as the unit of distances
we can find the Lorentz factors in dimensional $z$, and also estimate the efficiency
of the acceleration in the whole jet (which equals to the mean value of $\Gamma$
over $\mu=600$). For example,
at $z=100 R_\star/\lrz_j=7\times 10^{11}$~$(R_\star/10R_\odot)(100/\lrz_j)$~cm
the mean $\lrz$ is $\sim 200$ in the cold/magnetized case,
and the total efficiency $\sim 1/3$.

The last column of Fig.~\ref{raref_gammacont} corresponds to a
cold/magnetized case with smaller $\lrz_j$ and higher $\sigma_j$
(such that $\mu=\lrz_j (1+\sigma_j)$ remains the same as in the other cases).
It is interesting to note that, since $\lrz_j\sigma_j$ is approximately the same as
before, the dependence of $\lrz$ on $r$ remains the same, see equation~(\ref{gamma-r}).

Fig.~\ref{lraref_gammacont} shows two solutions with $\expov>1$, one cold/magnetized
(first column) and one hydrodynamic (second column).
The initial flow in not uniform now, with the density increasing as we move away
from the corner along constant $\theta$.
This allows for redistribution of the streamlines and acceleration even before
the head of the rarefaction is crossed. This is indeed seen in the figures, in both
the initial increase of the Lorentz factor as well as in the bending of the flow.
However, besides the initial phase the flows are very similar to the
corresponding rarefaction cases $\expov=1$.

\section{Summary and Discussion}\label{discuss}
In the present paper we develop a model for the steady-state, relativistic,
magnetohydrodynamic rarefaction wave. We use the method of self-similarity to
reduce the system of partial differential equations to ordinary ones, which we then
solve numerically.
The model is a generalization of existing works for unmagnetized and nonrelativistic
gas and can be applied in cases where plasma flows around a corner
(equivalently it loses its external support at some position).

We apply the model to long/soft GRB jets, which are formed inside the progenitor star
and lose their external support when they cross the stellar surface. In particular, we
used the model and successfully interpret the results of
recent numerical simulations that show a spurt of acceleration
in these jets, and more generally, whenever
a contact discontinuity with a relativistic flow along its plane is present.

Between models with the same energy-to-mass flux ratio we find that the {\em rarefaction
acceleration} is much faster in magnetized than in hydrodynamic flows.
Analytical scalings derived in Appendix~\ref{appendixb} helped to quantify this
behavior.
For the cold/magnetized case we find that the flow reaches $\lrz=\mu/2$
(half of its maximum value, i.e., 50\% efficiency of acceleration)
at distance
\begin{eqnarray}
r=\sigma_j \lrz_j |x_i|
=7\times 10^{11} \sigma_j
\left(\frac{|x_i|}{R_\star/\lrz_j}\right)
\left(\frac{R_\star}{10R_\odot}\right) \mbox{ \rm cm.}
\end{eqnarray}
The above rough estimation corresponds to $|x_i|=R_\star /\lrz_j$ being
the distance of the corner from the rotation axis, and $R_\star =10 R_\odot$
for the stellar radius.
(For the hydrodynamic case this distance is a few orders of magnitude larger.)

Our model assumes planar geometry and symmetry, which only locally hold near the
points where the surface of the jet intersects the stellar surface.
Improvements include axisymmetric studies, and also to take into account the
reflection of the wave on the rotation axis, which will possibly cause
the Lorentz factor to saturate at a value smaller than the maximum
(a crude approximation of the time needed for the information to
start from a fluid parcel passing the corner, hit the axis and
come back at the same parcel is $\sim 2 (R_\star /\lrz_j)/|\theta_h|
\sim 2 R_\star / \sigma_j^{1/2}$).
Axisymmetric studies are inherently nonuniform (one of the reasons
being that the magnetization vanishes on the rotation axis
where the azimuthal magnetic field should be zero). For this reason comparison
of numerical simulations of axisymmetric jets with
our model that assumes a uniform jet initially should be done with caution
at distances far away from the corner.

Another limitation is the assumption of a zero external pressure
outside the progenitor star. A finite external pressure will create
a standing shock and a contact discontinuity between the jet and its environment,
and also limit the terminal Lorentz factor to some value smaller than its maximum.
Our model cannot capture this inherently non-self-similar geometry. Nevertheless
it describes the basic physics of the mechanism and gives quantitatively correct
results for most of the rarefaction acceleration phase, till the point where the
shock is crossed. Since the pressure contrast inside and outside the progenitor star
is expected to be high, only the small shocked outflow part cannot be described
by our model.

All our findings are very similar to the ones discussed in
\cite{KVK10}. This is surprising at first, since their study is
time dependent and one dimensional in space, while ours is steady-state
and two dimensional in space. The reason for this similarity is the
so-called frozen pulse approximation, first introduced by \cite{PSN93}
for a relativistic hydrodynamic flow and extended by \cite{VK03a}
for the full relativistic MHD case.
According to this approximation, when a time dependent flow
is ultrarelativistic and superfast-magnetosonic, it can be described
using steady-state equations. The full mathematical proof can be found in
Appendix~\ref{appendixc}.
The physical reason is that each part of the flow moves practically with $c$
and cannot communicate with neighboring parts through fast-magnetosonic waves
(which also move at most with $c$). Thus, a possible time dependence
of the flow quantities at some point of space is carried with the flow
as a frozen pulse, and the motion of each part is effectively time independent.

\section*{Acknowledgments}
We thank the referee for many helpful comments.
This research has been co-financed by the European Union
(European Social Fund -- ESF) and Greek national funds through the
Operational Program ``Education and Lifelong Learning'' of the
National Strategic Reference Framework (NSRF) - Research Funding Program:
Heracleitus II. Investing in knowledge society through the European Social Fund.
NV acknowledges partial support by the Special Account for
Research Grants of the National and Kapodistrian University of Athens
(``Kapodistrias'' grant no 70/4/8829).


\appendix
\section{Fast Magnetosonic waves}
\label{appendixa}
Suppose that we study a magnetosonic disturbance on the poloidal plane.
Its phase speed in the comoving is
$\omega_{\rm co}/k_{\rm co}=\pm\cf$ with
\begin{eqnarray}
\cf=c \sqrt{\frac{\sigma(1+u_s^2)+u_s^2}{(1+u_s^2)(1+\sigma)}}
\label{omegaco}
\end{eqnarray}
(e.g., using the expressions given in Appendix C of \citealt{VK03a} for
propagation normal to the magnetic field,
$\bmath k_{\rm co} \bot \bmath B_{\rm co}$).
Here $\sigma = B_{\rm co}^2/4 \pi \enth \dens c^2 $
and $B_{\rm co}^2 = B^2-E^2 = B^2/\lrz^2$.
The corresponding proper speed (over $c$) is
\begin{eqnarray}
u_{\rm f}= \frac{\cf/c}{\sqrt{1-\cf^2/c^2}}
=\sqrt{\sigma (1+u_s^2) +u_s^2} \,.
\label{omegacoproper}\end{eqnarray}
Since the propagation is isotropic, the group velocity
is equal to the phase velocity, $v_{g\, {\rm co}} = c_{\rm f}$.

Transforming the dispersion relation to the central object's frame, we get
\begin{eqnarray}
\frac{\lrz\left(\omega - \bmath v \cdot \bmath k\right)}
{\sqrt{c^2 k^2-\omega^2}} =\pm u_{\rm f}  \,,
\end{eqnarray}
or equivalently
\begin{eqnarray}
\frac{\omega/k - \bmath v \cdot \bmath k / k}
{1-\omega \bmath v \cdot \bmath k/ k^2c^2} =\pm \frac{cu_{\rm f}}
{\sqrt{1+u_{\rm f}^2+(\lrz\bmath v\times\bmath k/ck)^2}}
\,.  \end{eqnarray}
The group velocity in the central object's frame can be found
from the transformation of the Lorentz factors
$\lrz_{g \, {\rm co}} = \lrz \lrz_{g}
\left(1-\bmath v \cdot \bmath v_g/ c^2 \right) $, or,
\begin{eqnarray}
\frac{1}{\sqrt{1-c_{\rm f}^2/c^2}} = \frac{\lrz
\left(1-vv_g \cos\thetag / c^2 \right) }{\sqrt{1-v_g^2/c^2}} \,,
\end{eqnarray}
where $\thetag$ is the angle between $\bmath v_g$
and the flow direction.
The above equation can be solved for $ v_g$:
\begin{eqnarray}\label{vglab}
v_g= \frac{\lrz^2 v\cos\thetag \pm \sqrt{u_f^2+1}
\sqrt{c^2 u_f^2-\lrz^2 v^2\sin^2 \thetag}}
{u_f^2+1+(\lrz v/c)^2\cos^2 \thetag} \,.
\end{eqnarray}
All directions $\thetag$ which give real values for the group velocity
form a Mach cone around the flow direction,
with half-opening $\tilde \thetag$
(the maximum allowed $\thetag$) given by
\begin{eqnarray}
\sin \tilde \thetag =\frac{u_{\rm f}}{\lrz v/c}
=\frac{\sqrt{\sigma (1+u_s^2) +u_s^2} }{\lrz v/c}
\,.
\label{mach_opening}
\end{eqnarray}
Note that the result is a direct generalization of the nonrelativistic
$\sin \tilde \thetag =c_{\rm f}/v$, with the proper speeds replacing
their Newtonian counterparts \citep{K80}.

An alternative way to find $\tilde \thetag$ follows:
\\
Assume a system of coordinates on the poloidal plane such that $\hat z$ is along
the flow velocity and $\hat x$ normal to it.
(Note that this is not the same with the $x-z$ system of coordinates
adopted in the main body of the paper, in which
the velocity makes an angle $\vvartheta$ with the $z$ axis.)
Consider a disturbance starting at $t=0$ from the line $x=z=0$
(for all $y$).
In the comoving frame the disturbance starts at $t_{\rm co}=0$
from the line $x_{\rm co}=z_{\rm co}=0$, and
after some time $t_{\rm co}>0$ affects a cylindrical regime
$z_{\rm co}^2+x_{\rm co}^2 = \cf^2 t_{\rm co}^2$,
since its group velocity is $v_{g\, {\rm co}} = c_{\rm f}$
(given by \ref{omegaco}).
In the central object's frame that regime is Lorentz transformed to
$\lrz^2 (z-vt)^2+x^2 = \cf^2 \lrz^2(t-vz/c^2)^2$, or
equivalently to the elliptic cylinder
\begin{equation}\label{front}
\frac{(\lrz^2+u_{\rm f}^2)^2}{(1+u_{\rm f}^2)u_{\rm f}^2 c^2 t^2}
\left(z-\frac{\lrz^2 vt}{\lrz^2+u_{\rm f}^2}\right)^2
+\frac{\lrz^2+u_{\rm f}^2}{u_{\rm f}^2c^2 t^2} x^2
=1 \,,
\end{equation}
an equation of the form ${\cal F}(x,z,t)=0$.
The area to which the disturbance is propagating is limited by the
envelope of these elliptic cylinders.
Solving the system ${\cal F}(x,z,t)=0 =
(\partial / \partial t) {\cal F}(x,z,t)$
we find the two characteristic planes $x/z=\pm
u_{\rm f}/\sqrt{\lrz ^2 -1-u_{\rm f}^2}$,
and thus the angle $\tilde \thetag$ between
the envelope and the flow velocity is given by
\begin{eqnarray}
\tan \tilde \thetag =
\frac{u_{\rm f}}{\sqrt{\lrz ^2 -1-u_{\rm f}^2}} \,,
\end{eqnarray}
an expression equivalent to \ref{mach_opening}.
\\
(The substitution of $x=v_g t \sin \thetag$ and $z=v_g t \cos \thetag$
in equation \ref{front} is an alternative way to find
equation~\ref{vglab} for the group velocity in each direction.)

\section{The MHD rarefaction wave}\label{appendixb}

Here we give the equations that characterize the
rarefaction regime $\theta_h\le \theta \le \theta_t$
(for the case $\expov=1$).

The head of the rarefaction corresponds to $\theta=\theta_h$.
Since $\theta_h=-\tilde \thetag $,
\begin{equation}
\theta_h = - \arcsin
\frac{u_{{\rm f}j}}{\lrz_j v_j/c}
\end{equation}
(using expression~\ref{mach_opening} of Appendix~\ref{appendixa}).
Here subscripts ``j'' refer to the uniform initial phase.

Using the normalized density $\varrho \equiv \dens/\dens_j$
as the independent variable,
equation~(\ref{x-M-v}) gives
\begin{eqnarray}
\enth=1+\left(\enth_j-1\right) \varrho^{\polind-1} \,, \quad
\sigma=\frac{\sigma_j \enth_j \varrho}
{1+\left(\enth_j-1\right) \varrho^{\polind-1}} \,,
\end{eqnarray}
equation~(\ref{ode2}) (which simplifies to ${\cal D}=0$) gives
\begin{eqnarray}
\ggf=\frac{1}{\enth \sigma u_{\rm f}}
\end{eqnarray}
with  $ u_{\rm f}=\sqrt{\displaystyle\frac
{\enth_j \sigma_j \varrho+\left(\polind-1\right)\left(\enth_j-1\right)
\varrho^{\polind-1}}{1+\left(2-\polind\right)\left(\enth_j-1\right)
\varrho^{\polind-1}}} $,
\\
equation~(\ref{odethe0}) gives $\vvartheta$ through
\begin{eqnarray}
\sin^2(\vvartheta-\theta)= \sin^2\tilde \thetag= \frac{u_{\rm f}^2}{\lrz^2-1}
\end{eqnarray}
with $\lrz =
\displaystyle\frac{\enth_j \lrz_j (1+\sigma_j)}{1+\left(\enth_j-1\right)\varrho^{\polind-1}
+\sigma_j \enth_j \varrho} $,
\\
and the differential equation~(\ref{ode10}) implies
\begin{eqnarray}
\theta=\theta_h+\int_\varrho^1
\frac{1}{\sqrt{\lrz^2 -1-u_{\rm f}^2}}
\frac{d \left(\varrho u_{\rm f}\right) }{d\varrho} \frac{d\varrho}{\varrho}  \,.
\end{eqnarray}
At the tail of the rarefaction wave $\varrho=0$
the thermal and magnetic energy fluxes vanish ($\enth=1$ and $\sigma=0$)
while $\ggf \rightarrow \infty$ and $\vvartheta = \theta$.
The position of the tail is $\theta=\theta_t$ with
\begin{eqnarray}
\theta_t=\theta_h+\int_0^1
\frac{1}{\sqrt{\lrz^2 -1-u_{\rm f}^2}}
\frac{d \left(\varrho u_{\rm f}\right) }{d\varrho} \frac{d\varrho}{\varrho}  \,.
\end{eqnarray}
The $x$ component of the velocity is $v_x=v_r \sin\theta +v_\theta \cos\theta $
with $v_\theta /c =u_{\rm f}/\lrz$ and $v_r /c =\sqrt{\lrz^2 -1-u_{\rm f}^2} \ /\lrz$.

For a highly superfast-magnetosonic
and ultrarelativistic flow $\lrz^2 \gg 1+u_{\rm f}^2 $
the relation $\theta-\varrho$ simplifies to
\begin{eqnarray}
\theta=-\frac{u_{{\rm f}j}}{\lrz_j}+\int_\varrho^1 \frac{1}{\lrz \varrho}
\frac{d \left(\varrho u_{\rm f}\right) }{d\varrho} d\varrho
=-\frac{u_{\rm f}}{\lrz} + \frac{v_x}{c} \,,
\end{eqnarray}
where  $v_x/c=$
\begin{eqnarray}
\frac{ \int_\varrho^1 \sqrt {\left[
\enth \sigma+\left(\polind-1\right)\left(\enth-1\right) \right]
\left[1+(2-\polind)(\enth-1)\right]} \frac{d\varrho}{\varrho}
}{\enth_j \lrz_j (1+\sigma_j)}
\,.
\end{eqnarray}

\subsection{The ultrarelativistic cold MHD limit}\label{coldapp}
In that limit $(\enth=1)$ the previous expressions can be greatly simplified.
We find $\sigma=\sigma_j \varrho$,
$\lrz=\lrz_j (1+\sigma_j)/(1+\sigma_j \varrho) $,
$u_{\rm f}=\sigma^{1/2}$,
$\ggf=\sigma^{-3/2}$,
and if the flow is highly superfast-magnetosonic
and ultrarelativistic $\lrz^2 \gg 1+\sigma$,
\begin{eqnarray}
\theta = \frac{2\sigma_j^{1/2}-3\sigma^{1/2}-\sigma^{3/2}}
{\lrz_j(1+\sigma_j)} \,.
\end{eqnarray}
For the head $(\sigma=\sigma_j)$ we get $\theta=\theta_h=-{\sigma_j^{1/2}}/{\lrz_j}$,
and for the tail $(\sigma =0)$ we find $ \theta=\theta_t=2|\theta_h|/(1+\sigma_j)$.
\\
The direction of the flow is given by
\begin{eqnarray}\label{B10}
\vvartheta = 2\frac{\sigma_j^{1/2}-\sigma^{1/2}}{\lrz_j(1+\sigma_j)} \,.
\end{eqnarray}
\\
The streamlines in the rarefaction regime $(\theta>\theta_h)$ are
(in polar coordinates) $r=r_\vflux \ggf = r_\vflux \sigma^{-3/2} $, or,
\begin{eqnarray}
\theta = \frac{2\sigma_j^{1/2}-3(r/r_\vflux)^{-1/3}-(r/r_\vflux)^{-1}}
{\lrz_j(1+\sigma_j)} \,.
\nonumber
\end{eqnarray}
Different values of $r_\vflux$ give different streamlines. For a streamline
that crosses the angle $\theta_h$ at $x=x_i$ we get $r_i=x_i/\theta_h$ and
$r_\vflux =r_i/\ggf_j = \sigma_j \lrz_j |x_i|$.

\subsection{The ultrarelativistic HD limit}
For the unmagnetized case ($\sigma=0$) similar approximations yield
\begin{eqnarray}\label{B11}
\enth=1+\left(\enth_j-1\right) \varrho^{\polind-1} \,, \quad
\lrz = \frac{\enth_j \lrz_j}{1+\left(\enth_j-1\right)\varrho^{\polind-1}} \,,
\end{eqnarray}
and if the flow is highly superfast-magnetosonic
and ultrarelativistic $\lrz^2 \gg 1+u_s^2 $
\begin{eqnarray}
\theta=\frac{-\enth u_s + \int_\varrho^1 \sqrt{
\left(\polind-1\right)\left(\enth-1\right)
\left[1+(2-\polind)(\enth-1)\right]} \frac{d\varrho}{\varrho}
}{\enth_j \lrz_j}
\quad \\
=\frac{
{\cal I} \left[\left(2-\polind\right)\left(\enth_j-1\right)\right]
-{\cal I} \left[\left(2-\polind\right)\left(\enth_j-1\right)
\varrho^{\polind-1}\right]}
{\enth_j \lrz_j \left(\polind-1\right)^{1/2}\left(2-\polind\right)^{1/2}}
-\frac{\enth u_s}{\enth_j \lrz_j}
\,, \nonumber
\end{eqnarray}
where
${\cal I} \left[\zeta\right] \equiv \zeta^{1/2} \left(1+\zeta\right)^{1/2}
+\ln \left[\zeta^{1/2}+\left(1+\zeta\right)^{1/2}\right] $.
\\
For the head $(\varrho=1)$ we get $\theta=\theta_h=-{u_{sj}}/{\lrz_j}$
and for the tail $(\varrho=0)$ we find $\theta=\theta_t=
\displaystyle\frac{
{\cal I} \left[\left(2-\polind\right)\left(\enth_j-1\right)\right] }
{\enth_j \lrz_j \left(\polind-1\right)^{1/2}\left(2-\polind\right)^{1/2}} \,.$
\\
For the distance form the corner we get
\begin{eqnarray}\label{B13}
\frac{r}{x_i/\theta_h}=
\frac{\ggf}{\ggf_j}=\varrho^{-(\polind+1)/2}
\sqrt{\frac{1+\left(2-\polind\right)\left(\enth_j-1\right)\varrho^{\polind-1}}
{\left(\polind-1\right)\left(\enth_j-1\right)}}\,.
\end{eqnarray}
\\ Note that for $\polind \rightarrow 2$, $\enth \rightarrow 1+\sigma$,
and $\enth_j \rightarrow 1+\sigma_j$
we recover the relations of Section~\ref{coldapp}
for the cold magnetized limit.
This is because, for a transverse magnetic field,
the magnetic pressure $B_{\rm co}^2/8\pi = (B^2/\lrz^2) / 8\pi$
is proportional to the square of the rest mass density
(see equation~[\ref{phipsi}]),
and thus it is analogous to a polytropic relation with index $\polind=2$.

\section{Comparison with the time-dependent rarefaction wave}
\label{appendixc}

For $v_y=0$, $\bmath B_p=0$, and $\partial/\partial y=0$,
the electric field is
\begin{equation}\label{ohmapp}
{\bmath{E}}= -\frac{v_x}{c}B \hat z + \frac{v_z}{c}B \hat x \,,
\end{equation}
and equations~(\ref{maxwell}--\ref{entropy}) become, after some manipulation,
\begin{equation}\label{continuityapp}
\frac{1}{c}\frac{\partial \left(\lrz \dens\right)}{\partial t}
+\frac{\partial}{\partial z} \left(\lrz \dens \frac{v_z}{c}\right)
+\frac{\partial}{\partial x} \left(\lrz \dens \frac{v_x}{c}\right) =0 \,,
\end{equation}
\begin{equation}\label{maxwellapp}
\left(\frac{1}{c}\frac{\partial}{\partial t}
+\frac{v_z}{c}\frac{\partial}{\partial z}
+\frac{v_x}{c}\frac{\partial}{\partial x} \right)
\left(\frac{B}{\lrz\dens}\right) =0 \,,
\end{equation}
\begin{equation}\label{entropyapp}
\left(\frac{1}{c}\frac{\partial}{\partial t}
+\frac{v_z}{c}\frac{\partial}{\partial z}
+\frac{v_x}{c}\frac{\partial}{\partial x} \right)
\left(\frac{p}{\dens^\polind}\right) =0 \,,
\end{equation}
\begin{eqnarray}\label{momentumappv}
\left(\frac{1}{c}\frac{\partial}{\partial t}
+\frac{v_z}{c}\frac{\partial}{\partial z}
+\frac{v_x}{c}\frac{\partial}{\partial x} \right)
\left(\enth \lrz+\frac{B^2}{4\pi \lrz \dens c^2}\right)
 \nonumber \\
=\frac{1}{\lrz \dens c^3}\frac{\partial}{\partial t}
\left(p+\frac{B^2}{8\pi \lrz^2}\right)
\end{eqnarray}
\begin{eqnarray}\label{transfieldappv}
\left(\lrz \enth+\frac{B^2}{4\pi \lrz \dens c^2}\right)
\lrz \dens v_z^2\left(\frac{1}{c}\frac{\partial}{\partial t}
+\frac{v_z}{c}\frac{\partial}{\partial z}
+\frac{v_x}{c}\frac{\partial}{\partial x} \right)
\left(\frac{v_x}{v_z}\right)
\nonumber \\
=\left(\frac{v_x}{c}\frac{\partial }{\partial z}
-\frac{v_z}{c}\frac{\partial }{\partial x}\right)
\left(p+\frac{B^2}{8\pi \lrz^2}\right)
\,.
\end{eqnarray}
(The last two equations correspond to the components of
the momentum equation along and normal to the flow.)
\\
For ultrarelativistic flows with $|v_x| \ll |v_z|$
we can simplify the above system, by
(i) using $v_z/c \approx 1$,
(ii) dropping the right-hand side of equation~\ref{momentumappv}
(since the left-hand side includes much larger terms
-- note that $dh = dP / \dens c^2$), and
(iii) noting that $|v_x \partial/\partial z | \ll
|v_z \partial/\partial x|$, which simplifies
equation~\ref{transfieldappv}.
Careful examination of equation~\ref{momentumappv} reveals that
the assumption $v_z\approx c$ holds only in the
superfast-magnetosonic regime\footnote{
In the part $\frac{v_z}{c} \left(1+\frac{B^2}
{4\pi \enth \lrz^2 \dens c^2}\right) \approx \sigma
\left(1-\frac{1}{2\lrz^2}\right) \left(1+\frac{1}{\sigma}\right)$
of that equation we kept the term $1/\sigma$ but not the
$1/\lrz^2$, something that is correct if
$\lrz^2 \gg \sigma$, or, $\lrz^2 \gg 1+u_{\rm f}^2$.  }.

The resulting system gives three integrals of motion
\begin{equation}
-\frac{B}{\lrz \dens c}=\velpot \,, \quad
\frac{p}{\dens^\polind }=Q \,, \quad
\enth \lrz+\frac{B^2}{4\pi \lrz \dens c^2} =  \mu \,,
\end{equation}
(which in principle are different for different parts of the flow),
and the equations
\begin{equation}\label{frozen1}
\left(\frac{1}{c}\frac{\partial}{\partial t}
+\frac{\partial}{\partial z}\right) \left(\lrz \dens
\frac{v_z}{c}\right)
+\frac{\partial}{\partial x} \left(\lrz \dens \frac{v_x}{c}\right) =0 \,,
\end{equation}
\begin{eqnarray}\label{frozen2}
\mu \lrz \dens c^2\left(\frac{1}{c}\frac{\partial}{\partial t}
+\frac{\partial}{\partial z}
+\frac{v_x}{c}\frac{\partial}{\partial x} \right) \frac{v_x}{v_z}
=-\frac{\partial }{\partial x}
\left(p+\frac{B^2}{8\pi \lrz^2}\right) .
\end{eqnarray}
By inspection of the previous equations we see that the
derivatives $\partial/\partial t$ and $\partial/\partial z$
always come as a combination $\partial/c\partial t
+\partial/\partial z$,
and thus the variables $z$ and $ct$ can be interchanged.
The steady state problem where $\partial/\partial t=0$
and the flow depends on $z$ and $x$ is
mathematically equivalent to the time-dependent
one-dimensional problem where $\partial/\partial z=0$
and the flow depends on $ct $ and $x$.

The above is a manifestation of the ``frozen pulse'' behavior
of an ultrarelativistic flow, first introduced by \cite{PSN93}
for hydrodynamic flows and extended by \cite{VK03a} in the MHD case.
Due to the ultrarelativistic and superfast-magnetosonic velocity
of the flow, any possible disturbance is traveling with it
and cannot affect the neighboring parts.
As a result the evolution of each fluid parcel
is essentially steady-state.
In fact, changing variables from $(x,z,t)$ to
$(x,z,s)$ where $s\equiv ct-z$, we transform
equations~\ref{frozen1}, \ref{frozen2} to
\begin{equation}\label{frozen1s}
\frac{\partial}{\partial z}\left(\lrz \dens \frac{v_z}{c}\right)
+\frac{\partial}{\partial x} \left(\lrz \dens \frac{v_x}{c}\right) =0 \,,
\end{equation}
\begin{eqnarray}\label{frozen2s}
\mu \lrz \dens c^2\left(\frac{\partial}{\partial z}
+\frac{v_x}{c}\frac{\partial}{\partial x} \right) \frac{v_x}{v_z}
=-\frac{\partial }{\partial x}
\left(p+\frac{B^2}{8\pi \lrz^2}\right) \,,
\end{eqnarray}
which are the same with the steady-state equations
in the same (ultrarelativistic) limit.
Note however that the partial derivatives
$\partial/\partial z$, $\partial/\partial x$
are now taken keeping $s$ (and not $t$) constant.

Since the motion is relativistic in the $z$ direction
the variable $s$ is practically constant for each fluid parcel
and corresponds to the time in which it passed a certain
position $z_i$. Without loss of generality we can
set $z_i=0$; in that case $t_i=s/c$.
The absence of $s$ and $\partial/\partial s$ in
equations~\ref{frozen1s},\ref{frozen2s}
means that they do not constrain the $s$ dependence on any flow
quantity ${\cal F}(x,z,s)={\cal F}(x,z,ct-z)$. This
dependence is determined by the initial/boundary conditions only,
i.e., by the values of the flow quantities for each fluid parcel
at time $t_i$ when it passes $z=0$.
In other words, we can find the evolution of a time-dependent
flow by applying steady-state solutions to each part of the flow
passing from $z=0$ at time $t_i=s/c$,
by changing only the boundary conditions
(see an example in Section~4.1.1 of \citealp{VK03a}).

In the particular case of the relativistic rarefaction wave, the
frozen pulse approximation obviously holds\footnote{
It can be easily checked that the requirement
$|v_x \partial/\partial z | \ll |v_z \partial/\partial x|$
indeed holds in the $\expov = 1$ case in which the
flow depends only on $z/x$.}.
As a result, the steady-state solutions considered in this work
can be used for the description of a time-dependent flow,
and this can be achieved by simply writing the similarity variable
as $z/x=(ct-s)/x = c(t-t_i)/x $.
Thus, we only need to make the substitution $z\rightarrow c(t-t_i)$
(with constant $t_i$ for each part of the flow)
in order to recover the equations of the time-dependent
rarefaction wave with ultrarelativistic velocity
in the $z$ direction, considered in \cite{KVK10}.
The worldlines of all fluid parcels passing at time
$t_i=0$ from the plane $z=0$ (at various $x<0$)
are equivalent to the streamlines of the steady-state model.
This is indeed the case in the numerical results.
Choosing initial conditions as the ones in Fig.~4 of \cite{KVK10}
we get practically identical results, by just
substituting $z\leftrightarrow ct$.

\label{lastpage}
\end{document}